\documentclass[useAMS,usegraphicx,usenatbib]{mn2e}
\usepackage{txfonts,amssymb,hyperref,subfigure,epsfig}

\def \aap{A\&A}

\def \aj{AJ}

\def \apj{ApJ}

\def \mnras{MNRAS}

\def \nat{Nat}

\newcommand{\Msun}{M_{\odot}}
\newcommand{\Mearth}{M_{\oplus}}
\newcommand{\Mstar}{M_{\star}}
\newcommand{\fracbrac}[2]{\left(\frac{#1}{#2}\right)}

\newcommand{\Fig}[1]{Fig \ref{#1}}
\newcommand{\Eqn}[1]{Eqn \ref{#1}}
\newcommand{\beq}{\begin{eqnarray}}
\newcommand{\eeq}{\end{eqnarray}}

\defcitealias{Thebault08a}{TMS08a}
\defcitealias{Thebault09}{TMS09}

\title[Terrestrial Embryo Migration around Binaries]{Outward Migration of Terrestrial Embryos in Binary Systems} 
\author[M.~J.~Payne, M.~C.~Wyatt \& P.~Th\'{e}bault]{
  Matthew~J.~Payne$^{1}$
  \thanks{E-mail:~\href{mailto:mpayne@ast.cam.ac.uk}{mpayne@ast.cam.ac.uk}(MJP)}, Mark~C.~Wyatt$^{1}$ and Philippe~Th\'{e}bault$^{2,3}$\\
  $^1$Institute of Astronomy, University of Cambridge, Madingley Road, Cambridge CB3 0HA, matthewjohnpayne@gmail.com\\
  $^{2}$Observatoire de Paris, Section de Meudon, F-92195 Meudon Principal Cedex, France\\
  $^{3}$Stockholm Observatory, Albanova Universitetcentrum, SE-10691 Stockholm, Sweden
}

\begin{document}

\date{Accepted -------. Received -------; in original form -------}
\pubyear{2009}

\maketitle
\label{firstpage}
\begin{abstract}
We consider the formation and migration of protoplanetary embryos in disks around the stars in tight binary systems (separations $\sim 20\,AU$). In such systems, the initial stages of runaway embryo formation are expected to only take place within some critical disk radius $a_{crit}$, due to the perturbing effect of the binary companions \citep{Thebault09}. We perform n-body simulations of the evolution of such a population of inner-disk embryos surrounded by an outer-disk of smaller planetesimals. Taking Alpha Centauri-B as our fiducial reference example in which $a_{crit} \sim 0.7\,AU$, and using a Minimum Mass Nebular Model with $\Sigma \propto a^{-3/2}$, we find that within $10^6 yrs$ ($10^7 yrs$), systems will on average contain embryos which have migrated out to $0.9\,AU$ ($1.2\,AU$), with the average outer-most body having a mass of $0.2\Mearth$ ($0.4\Mearth$). Changes to increase the surface density of solids or to use a flatter profile both produce increased embryo migration and growth. At a given time, the \emph{relative} change in semi-major axis of the outer-most embryo in these simulations is found to be essentially independent of $a_{crit}$, and we note that little further embryo migration takes place beyond $10^7\,yrs$. We conclude that the suppression of runaway growth outside $a_{crit}$ does \emph{not} mean that the habitable zones in such tight binary systems will be devoid of detectable, terrestrial mass planets, even if $a_{crit}$ lies significantly interior to the inner edge of the habitable zone.
\end{abstract}

\section{Introduction}\label{Intro}
The majority of stars occur in multiple systems \citep{Eggleton08} and $\sim20\%$ of detected exo-planets have been found in binary systems \citep{Desidera07}, hence the importance of studying planet formation in such systems. Most of these systems have very large separations for which the influence of the companion star on the planet region is probably limited. However, 3 planets have been detected in tighter binaries whose separation, $a_b$, is around $20\,AU$, with one such system ($\gamma$ Cephei) having a planet with a minimum mass of $1.6M_{Jup}$ orbiting at $2.1\,AU$ \cite{Hatzes03H}. While such systems might harbour planetary orbits which have a long term stability in regions up to 3 to $5\,AU$ from the primary \citep[E.g.][]{Holman99}, they are thought to present much more challenging environments for planet \emph{formation} in these regions (E.g. \citet{Kley07} and references therein). 

In recent years, this issue of planet formation in tight binaries \footnote{We refer here to "tight" binaries as systems with separations around 20\,AU. Systems with even smaller separations are left out of this study since they usually imply unstable orbits in most of the potential terrestrial planet region and are thus of limited interest for the present problem} has been investigated in numerous studies. They have shown that the late stages of planetary accretion, i.e. the final assembly of large Lunar-sized embryos, can proceed in a domain almost as large as the orbital stability region \citep{Barbieri02, Quintana07, Guedes08}. However, the earlier stages, starting from kilometre-sized planetesimals and leading through runaway and oligarchic growth to the embryos, are much more affected by the binary companion. Indeed, the coupled effects of secular perturbations from the companion star and gas friction lead to strong orbital phasing as a function of planetesimal sizes. Such phasing results in high impact velocities between differently sized objects, causing accretion to be significantly slowed or even turned to fragmentation for a large range of parameter space. As a result, only the innermost regions of the protoplanetary disk, beyond a critical distance $a_{crit}$ from the central star, might allow the in-situ formation of planets \citep[E.g.][]{Thebault06,Paardekooper08a,Xie08}. For binaries with $a_b \sim 20\,$AU, $a_{crit}$ could be (depending on the system's eccentricity and mass ratio) well inside 1\,AU \citep[see for example Fig.8 of ][]{Thebault06}. This could pose serious problems for explaining the presence of a planet like that of the $\gamma$ Cephei system at 2.1\,AU \citep[see for instance the numerical investigations of][]{Paardekooper08a,Xie08}, or that at 2.6\,AU in HD196885 \citep[although the orbit of the binary is here not fully constrained, see][]{Correia08a}.

A good illustration of these problematic issues is the case of our closest stellar neighbour, the tight binary Alpha Centauri ($\alpha$ Cen). To date, no planets have been found in the system (\citet{Endl01} rule out any planets with masses $>2.5M_{Jup}$), but a recent study by \citet{Guedes08} suggests that, \emph{if} terrestrial planets do exist in this system, then $\alpha$ Cen-B would be the ideal candidate for the detection of potential terrestrial planets via the radial velocity technique, due to the proximity and quiet nature of the star. However, despite long-term stability as well as embryo-assembly studies which have given very promising results for the regions interior to $2\,AU$ \citep{Holman97, Barbieri02, Quintana02, Guedes08}, \citet{Thebault08a,Thebault09} have shown that the $a_{crit}$ for planetesimal accretion could be located as close as 0.5\,AU from each star. This would place it at the inner edge of the habitable zone (HZ) of $\alpha$ Cen-B \citep[$\sim0.5-0.9$\,AU, see][]{Guedes08}, but well inside that of $\alpha$ Cen-A \citep[$\sim1.0-1.3$\,AU, see][]{Barbieri02}. Although a recent study by \citet{Xie09} shows that a small inclination between the planetesimal disc and the binary orbital plane would help push the critical radius further out, the \emph{in-situ} formation of planets in both HZs remains highly problematic. However, \citet{Thebault09} point out that the \emph{presence} of planets in regions beyond $a_{crit}$ cannot be ruled out if additional mechanisms are invoked. Such mechanisms include (i) a change in dynamics after gas dispersal, (ii) a change in the binary orbit properties at some early point in the history of $\alpha$ Cen, or (iii) outward migration of the planetary embryos which formed in regions with $a<0.5\,AU$. \citet{Thebault09} go on to discuss points (i) and (ii) in some detail, ruling out the former and acknowledging the latter as a possible solution, but leave the detailed study of (iii) to future investigations.

Our aim in this paper is to address this crucial issue in more details. The main question we want to answer is the following: could large embryos, accreted in the inner regions interior to $a_{crit}$, later migrate outward and end up beyond $a_{crit}$? And if so, could this radial displacement be enough to place planets in the HZs of tight binary systems? 

We note that the migration of planetary embryos has been observed in many previous investigations. Among the many examples of this, \citet{Chambers01} demonstrate mixing and rearrangement of embryos due to embryo-embryo scattering, whilst \citet{Kirsh09} show embryo migration which is driven by embryo-planetesimal interactions. We would thus very generally expect to observe embryo migration within our simulations and simply wish to understand the degree and nature of the phenomenon.

We numerically investigate this problem, using an N-body code, taking the $\alpha$ Cen system as a typical reference example of a tight binary.

\section{Model and Method}\label{Method}
\subsection{The Disk}
We start by considering the protoplanetary disk around one star in a binary system, just after the dissipation of the disk gas. The disk is thus taken to consist solely of solid material distributed with a surface density profile $\Sigma = \Sigma_0 a^{-\alpha}$, where $\Sigma_0$ is the normalised surface density at $1\,AU$ (in $\textrm{g cm}^-2$), $a$ is the semi-major axis (in $\,AU$) and $\alpha$ is a positive constant. Unless otherwise stated, we take $\Sigma_0 = 10$ and $\alpha = 3/2$ as used in the nominal Minimum Mass Nebular Model (MMNM, \citet{Hayashi81}) case of \citet{Thebault09}.

We divide the disk into two zones, (i) an inner zone, $a < a_{crit}$, where runaway growth is assumed to have taken place and embryos have grown to isolation, and (ii) an outer zone, $a > a_{crit}$, where runaway growth was suppressed and the material in the disk is composed of smaller mass planetesimals. We take $\alpha$ Cen-B as our reference case in the majority of our simulations, taking $a_{crit} = 0.7\,AU$ for such a system. We then set inner and outer simulation boundaries at 0.2 and $1.5\,AU$ respectively, meaning that the total mass of embryos in the inner zone is $\sim 2\Mearth$, while the total mass of the outer zone is also $\sim 2 \Mearth$ (the width of the outer region was chosen to give a total mass approximately equal to that of the inner zone when $a_{crit} = 0.7\,AU$).

\subsection{The Inner Zone}
We expect that in the regions of the disk interior to $0.7\,AU$, where runaway growth can take place, growth to isolation will have occurred prior to the gas disk dissipation, resulting in a chain of isolated embryos. 

To check the validity of this assumption, consider the growth timescale of a protoplanetary embryo of mass $M$ embedded in a population of planetesimals of mass $m$ in the \emph{absence} of an external perturber: this is \citep{Kokubo02}:
\begin{eqnarray}
  \tau &=& 1.2\times10^5 \fracbrac{\Sigma_0}{10\textrm{g cm}^{-2}}^{-1} \fracbrac{M}{\Mearth}^{1/3} \fracbrac{a}{1\textrm{ AU}}^{3/5} \fracbrac{\Mstar}{\Msun}^{-1/6} \nonumber \\ 
       && \times \fracbrac{\Sigma_{Gas,0}}{2400\textrm{g cm}^{-2}}^{-2/5} \fracbrac{m}{10^{18}g}^{2/15} yrs. \label{EQN:isolation}
\end{eqnarray}
where $\Sigma_{Gas,0}$ is the surface density of gas in the disk around the star of mass $\Mstar$ and $K$ is an integer used to specify the width of the annular feeding zone, $\Delta a = Kr_H$, in terms of the Hill radius, $r_H = a (M/3\Mstar)^{1/3}$.
Such an embryo would thus grow to an isolation mass 
\begin{eqnarray}
  M_{iso}        &=& \left(2\pi K \Sigma_0\right)^{3/2} \left(3\Mstar\right)^{-1/2}\fracbrac{a}{1\textrm{ AU}}^{3/4}\left(AU\right)^{3}  \label{EQN:mass}
\end{eqnarray}
in a time, $t\sim 3\tau$, where the factor of 3 arises on integrating $\tau \propto M^{1/3}$.

This calculation ignores the refinements of the \citet{Thebault09} work on the suppression of runaway growth in the presence of a perturber. Critically, Eqn \ref{EQN:isolation} requires that embryos and planetesimals of mass $\sim 10^{18}g$ ($\sim5km$) have already formed and continue to exist. Despite these obvious drawbacks, we employ Eqn \ref{EQN:isolation} to give us an approximate estimate of the timescale for growth to isolation, and suggest that because (a) $\tau$ is less than $10^5yrs$ at small semi-major axes, (b) a putative gas disk around Alpha Cen is likely to have dispersed on a $> 10^6$ year timescale and (c) the suppression of the growth rate in the inner region is unlikely to be more than an order of magnitude, it thus seems likely that growth to isolation will have occurred in the region interior to $0.7\,AU$ before the gas disk dissipates. Hence we feel it is reasonable to assume that there exists a chain of isolated embryos at $a < a_{crit}$. It is unlikely that any significant evolution beyond this stage would have taken place at the time of disk dispersion, due to the very long orbit-crossing timescale in the presence of gas damping.

For $K=7$, $\Sigma_0 = 10\,g\,cm^{-2}$ and $a\sim 0.5\,AU$, the isolation mass will be $M_{iso}\sim 2\times 10^{26}g $ and the isolation separation $\Delta a_{iso} \sim 0.02\,AU$, meaning that $e < 0.02$ for non-overlapping orbits. As such, a reasonable set of initial conditions for the inner regions of our simulations seems likely to be a chain of isolation mass embryos, around 50 in number, with average eccentricities $<e>\sim 0.01$.

We simulate these isolation mass embryos as a population of fully interacting massive particles within the n-body code, i.e. the embryos interact gravitationally both with themselves (embryo-embryo) and with the smaller planetesimals (embryo-planetesimal) which occupy the outer zone. Note that in the majority of the simulations there are \emph{no} planetesimals within the inner zone.

\subsection{The Outer Zone}
The mass of solid material in the outer disk ($\sim 2\Mearth$) is divided into a population of equal mass planetesimals, which are randomly distributed radially according to the $a^{-3/2}$ profile, and the majority of the simulations use 100 planetesimals in the outer zone, meaning that each has a mass $m \sim 0.02\Mearth$.

The planetesimals are modelled within the n-body code as test particles with an associated mass (massive test particles), meaning that they interact gravitationally with the embryos in the system, but they do \emph{not} undergo planetesimal-planetesimal scattering amongst themselves. Some details of the effect of this approximation are given in \S \ref{Approximations}.

\subsection{Numerical Method}
We use the {\sc mercury} n-body package of \citet{Chambers99} to perform simulations of the evolution of such a protoplanetary disk with the aim of understanding the degree to which embryos can migrate from the inner zone ($a < a_{crit}$) to the outer zone ($a > a_{crit}$).

A summary of all the simulations performed is given in Table \ref{TABLE:Sims}. Unless otherwise detailed in that table, the simulations were set up and performed as follows: 
\begin{itemize}
\item As the gas disk is assumed to have dissipated prior to the start of our simulations we do \emph{not} impose gas drag / damping on any of the simulated embryos or planetesimals.
\item The solid disk is split into two regions: (i) The \emph{inner} region, $0.2 < a < a_{crit}$, populated with isolation-mass embryos, and (ii) The \emph{outer} region, $a_{crit} < a < 1.5$, populated with a swarm of lower-mass planetesimals. We take $a_{crit} = 0.7\,AU$ in the majority of our simulations, 
\item The majority of the simulations are based around a single star and do \emph{not} include the binary perturber in the n-body simulation. We note that, perhaps surprisingly, excluding the binary perturber from the simulations does \emph{not} strongly affect the evolution in the semi-major axis, mass or eccentricity of the outward-migrating embryos and hence we omit the binary perturber from the majority of our simulations. We delay a more detailed examination and discussion of this result until \S \ref{Binary Companion}.
\item We run {\sc mercury} in ``hybrid'' mode for almost all simulations, allowing rapid simulation via symplectic integrator when there are no close encounters, but also allowing accurate resolution of any close encounters (by using the Bulirsch-Stoer integrator). However, for any simulations which include the binary companion, we solely use the Bulirsch-Stoer integrator.
\item Collisions occur when two bodies approach to within the sum of their radii, with the radii being calculated assuming a density, $\rho = 3 \textrm{ g cm}^{-3}$. Any collisions are assumed to result in perfect mergers.
\item For any given parameter set we conduct 10 version of each simulation, keeping the number of embryos and planetesimals constant across all 10 simulations, but randomising the orbital angular elements of these bodies. The results of such an ensemble of simulations allows some insight into the variations inherent in the chaotic evolution of the system.
\item In our fiducial simulation (set A), we have $\sim 56$ embryos in the inner zone and $100$ planetesimals in the outer zone. Generally, we have kept these numbers constant in the other simulation sets. However, variations in other parameters (E.g. surface density) can indirectly result in the initial number of embryos varying (as a result of Eqn. \ref{EQN:mass}): where this occurs we note the change and also conduct additional simulations to check that the change in embryo number is not affecting the results. More details on the numbers of embryos and planetesimals in each simulation set are included in the pertinent sections.
\end{itemize}

\section{Results}\label{Results}
We provide a brief summary of the simulations performed and the key results obtained in Table \ref{TABLE:Sims}.
\begin{table*}
\begin{minipage}{1.0\textwidth}
\caption{List of Simulations Performed. Note that $\Sigma_E$ refers to the surface density of embryos, whilst $\Sigma_P$ refers to the surface density of planetesimals. The migration of the outermost embryo is expressed via the quantity $\eta = (a_{outer-final} - a_{outer,initial})/a_{outer,initial}$.}
\label{TABLE:Sims}
\begin{tabular}{lp{4cm}cp{8cm}}
\hline
\multicolumn{1}{|c|}{Set} & 
\multicolumn{1}{|c|}{Parameter Variation} & 
\multicolumn{1}{|c|}{Domain of Variation} & 
\multicolumn{1}{|c|}{Key Results} \\
\hline
Set A  &    Fiducial MMNM simulation                                                                                  &                -                  &       Outer most embryo migrates to $0.9\,AU$ ($\eta = 0.3$) within $10^6yrs$ and has a mass of $0.2\Mearth$. 
\\
Set B       &    Length of time over which simulations are conducted, $T$                                                  &      $10^6 - 10^8$ yrs            &       As we lengthen the simulations by a factor of 10 the relative change in semi-major axis more than doubles ($\eta (t=10^6)=0.3$, $\eta (t=10^7)=0.8$) and masses almost double. Little outward migration is observed between $10^7$ and $10^8$ years.
\\
Set C       &    Surface density normalisation of (fully interacting) embryos in the inner zone, $\Sigma_{E,Inner}$        &      $5 - 50\textrm{g cm}^{-2}$   &       Increased surface densities increase outward migration and cause order-of-magnitude increases in masses of the outer bodies.
\\
            &       and ratio of outer surface density to inner surface density, $\Sigma_{P,Inner}/ \Sigma_{E,Inner}$       &      $0 - 3$                     &       Higher ratios of planetesimal to embryo surface density cause increased migration and order-of-magnitude increases in masses of the outer bodies
\\
Set D       &    Surface density profile of the disk, $\alpha$                                                             &      $-1.5 \rightarrow 0$         &       Shallower surface density profiles cause increased migration due to the relatively larger amount of exterior mass, with $\eta=1.0$ at $10^6\,yrs$ for $\alpha = 0$.
\\
Set E       &    Position of the Inner-Outer Boundary, $a_{crit}$                                                          &      $0.4 - 1.0\,AU$              &       There is a large scatter in the degree of relative migration, but are approximately consistent with $\eta$ being independent of $a_{crit}$.
\\
Set F       &    \emph{Number} of (fully interacting) embryos in the inner zone, $N_{E,Inner}$                             &       $0 - 100$                   &       Both the amount of outward scatter and the final mass of the migratory planet are strongly dependent on $N_{E,Inner}$.
\\
            &       and  \emph{Eccentricity} of (fully interacting) embryos in the inner zone  $e_{E,Inner}$               &       $0.01 - 0.5$                &       The initial eccentricity is unimportant for the migration as long as $e_{E,inner} < 0.5$, but does play a role in determining final masses.
\\
Set G       &    Fraction of the inner disk which is composed of embryos, $f$                                              &      $0.1 - 1$                    &       If the total surface density is held constant, then lower $f$ results in slightly more migration, but slightly less massive planets.
\\          &                                                                                                              &                                   &       If the total surface density increases at constant density of embryos, then both outward migration and mass are increased.  
\\
Set H       &    Number of (massive test-particle) planetesimals in the outer zone, $N_{P,Test,Outer}$                     &       $100 - 1,000$               &       When the \emph{individual} mass of the planetesimals is reduced, an insignificant decrease in the amount of migration is observed.
\\
Set I       &    Fully interacting planetesimals in the outer zone, $N_{P,Full,Outer}$                                     &       $100 - 250$                 &       The use of massive test particles as opposed to fully-interacting particles for the planetesimals does not strongly change the degree of migration. 
\\
Set J       &    Addition of binary perturber to the simulated systems                                                     &      -                            &       The addition of a binary perturber leaves migration distance almost identical and slightly decreases mass growth.
\\
\hline
\end{tabular}
\end{minipage}
\end{table*}

\subsection{Typical Results}\label{Typical}
\begin{figure}
\centerline{\psfig{figure=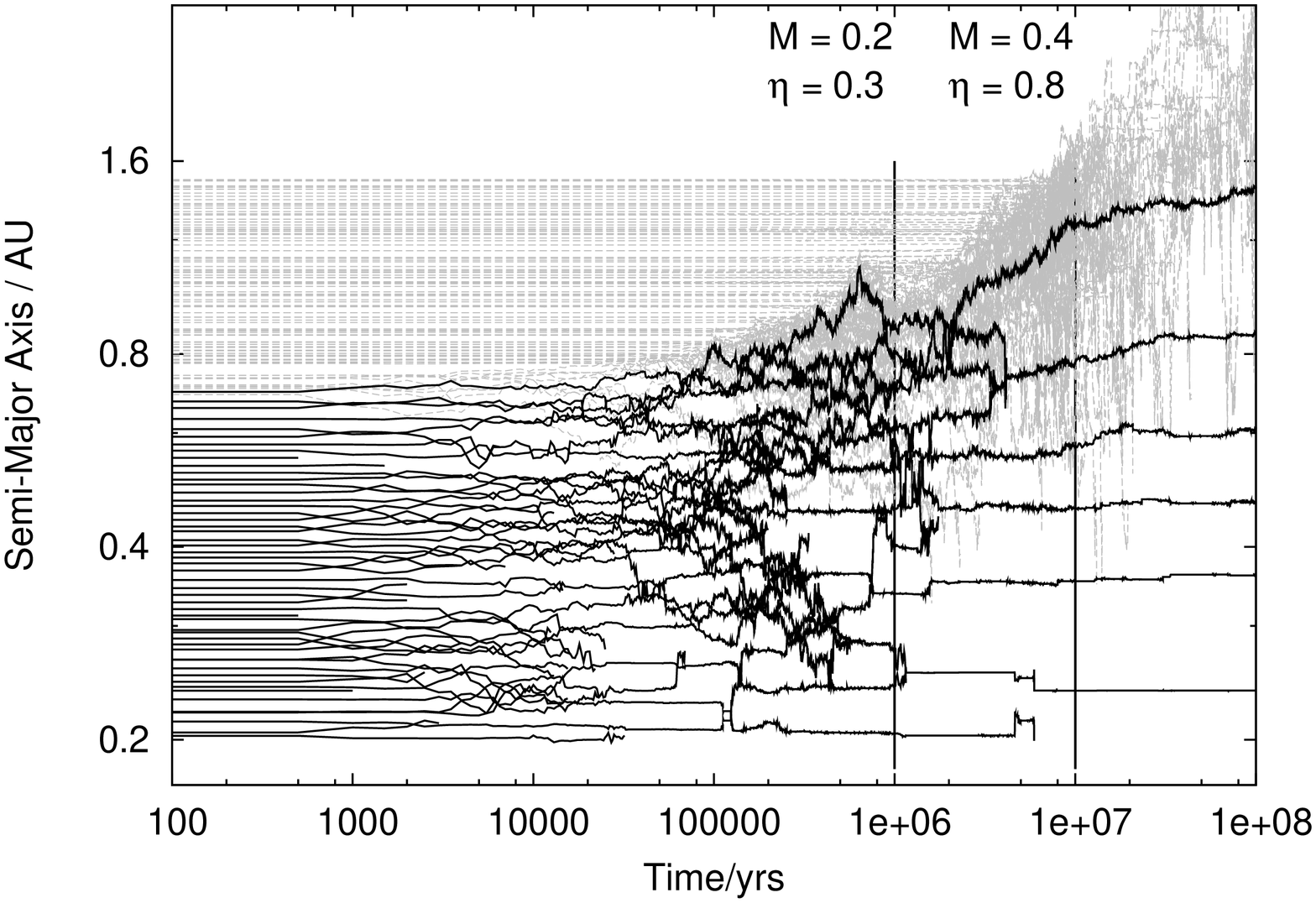,angle=-90,width=\columnwidth}}
\centerline{\psfig{figure=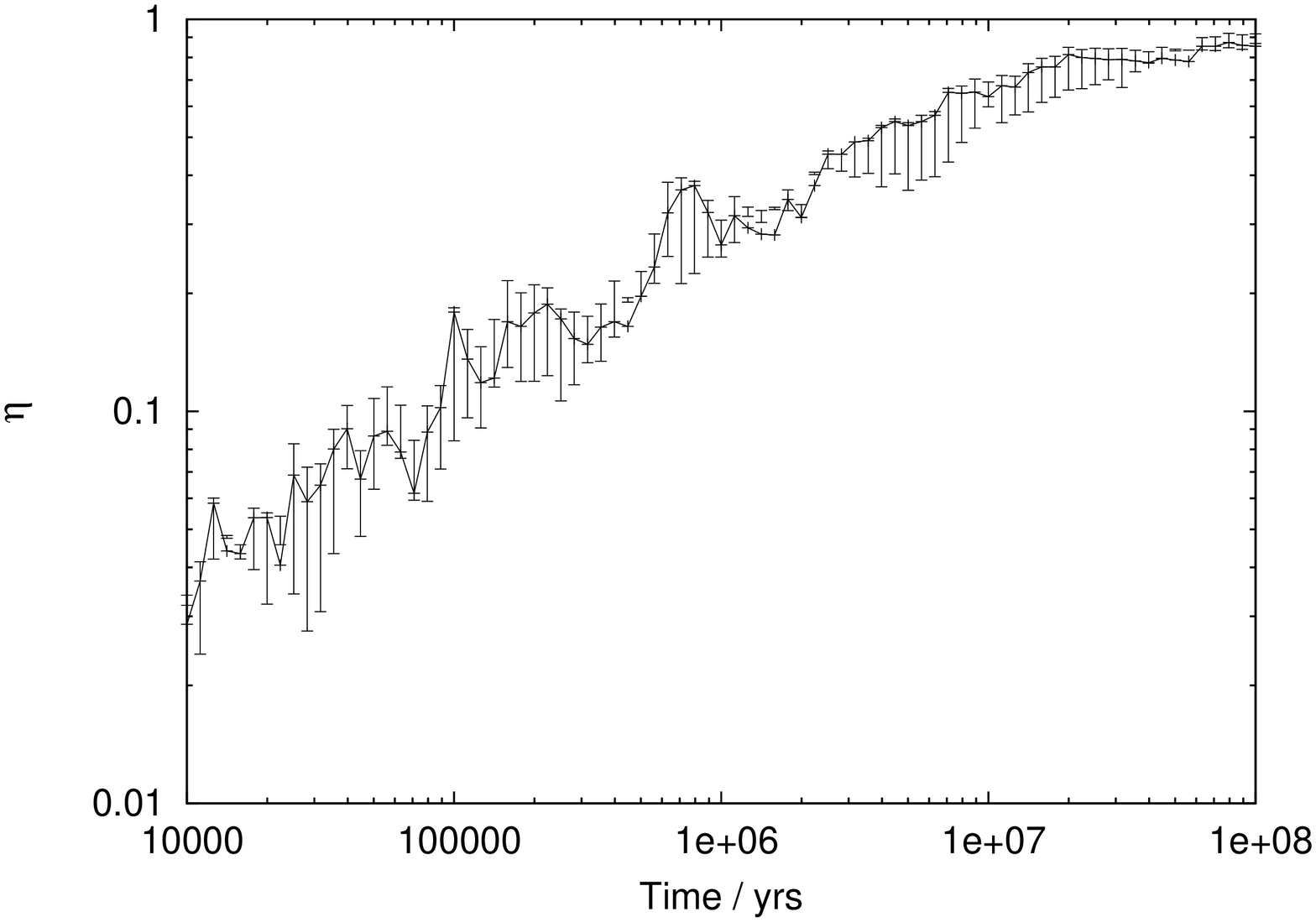,angle=-90,width=\columnwidth}}
\caption{Top: Semi-major axis evolution in an example of the fiducial MMNM simulation. The embryos which initially have $a<0.7\,AU$ are plotted in black, the planetesimals which are initially in the outer region are grey. For the outer-most embryo the masses (in units of $\Mearth$) and relative change in semi-major axes ($\eta = (a_{outer,final} - a_{outer,initial})/a_{outer,initial}$) are labeled at $10^6yrs$ \& $10^7yrs$. We see that significant embryo migration out to $\sim 1\,AU$ is observed. Bottom: The value of $\eta$ as a function of time, averaged over 10 simulations, plotting the median value, as well as the lower and upper quartile bounds.}
  \label{FIG:Fiducial}
\end{figure}

Consider a MMNM disk in which isolation mass embryos are initially confined to the inner region ($0.2 - 0.7\,AU$), while the outer region ($0.7 - 1.5\,AU$) has had runaway growth suppressed and is populated by smaller mass planetesimals (simulated using massive test particles). We conduct a set of 10 simulations of such a disk, assigning to each simulation a different set of randomised angular orbital elements. We label this fiducial set, A, and display in Fig \ref{FIG:Fiducial} the full orbital evolution of all of the particles in one of the simulations. 

We find that at $10^6yrs$, the initial 56 embryos have reduced (via collision) to $\sim 10$ in number, with the outer body having migrated outwards to $\sim 0.9\,AU$. If we define the relative change in semi-major axis,
\begin{eqnarray}
\eta &=& \frac{a_{outer,final} - a_{outer,initial}}{a_{outer,initial}}, \label{EQN_ETA_DEFN}
\end{eqnarray}
then we find that $\eta = 0.29_{-0.09}^{+0.08}$, where the errors indicate the variations to the upper and lower quartiles. The mass of this outer-most body has grown from 0.06 to $0.2\Mearth$, primarily through the accretion of planetesimals. 

Continuing our simulations of set A onto $10^7yrs$ (set B and Fig \ref{FIG:Fiducial}), we find that on average: (i) Further collisions have taken place, reducing the number of embryos within $1.5\,AU$ to $\sim 6$ and (ii) The outer-most embryo has now migrated to $\sim 1.25\,AU$ (giving $\eta \sim 0.8_{-0.24}^{+0.06}$), while its mass has further doubled to $\sim 0.4\Mearth$.

Only a very small amount of additional migration is observed between $10^7$ and $10^8\,yrs$, suggesting that the degree of migration observed by $10^7\,yrs$ suffices to give a reasonable approximation to the overall degree of migration expected. We should emphasise though that by this stage of the simulations the outer-most embryo has migrated to a position rather close to the outer edge of the simulated planetesimal zone at $1.5\,AU$, so any future simulations which wanted to properly examine migration to $10^8\,yrs$ and beyond, would have to be run with many more planetesimal bodies spread out to much larger semi-major axes.

We thus find that the embryos which start with semi-major axes $a < a_{crit}$ will migrate outwards over time as they interact with the exterior planetesimals. This means that despite the early suppression of runaway growth in regions beyond $a_{crit}$, the Habitable Zone (HZ) of the disk (which would initially have been devoid of large bodies if $a_{crit}$ lay at a smaller semi-major axis than the inner edge of the HZ) can become populated with large protoplanetary embryos within the relatively short time of $10^7$ years.

We note that the masses of these outer bodies in our simulations remain rather small compared to the detection limits of current technology such as Kepler \citep[see E.g.][]{Deeg02}), but we note that at the $10^7 - 10^8$ year point in our simulations, the systems are still rather densely packed (6 bodies within $1.5\,AU$). We expect that extended simulations (to $10^9$ years) would suffer further collisional evolution, yielding systems with rather fewer but more massive bodies, bringing many within detectable limits.

%

\subsection{Variation in Surface Density}\label{Results_Sigma}
We now look at the results for set C in which we vary the surface density of the inner zone between $5$ and $50 \textrm{g cm}^{-2}$, and also look at the effect of varying the surface density ratio between the inner and outer zones (i.e. a scenario in which the inner zone is depleted compared to the outer region, or a scenario in which the outer region has become enriched). Note that the simulations with surface density ratios of 2 or 3 were extended outwards so that the outer edge of the simulation zone was at 3.5AU, rather than $1.5\,AU$, to allow for the possibility of significantly greater outward migration (the number and total mass of the planetesimals was increased according to the surface density profile). 

\begin{figure}
\centerline{\psfig{figure=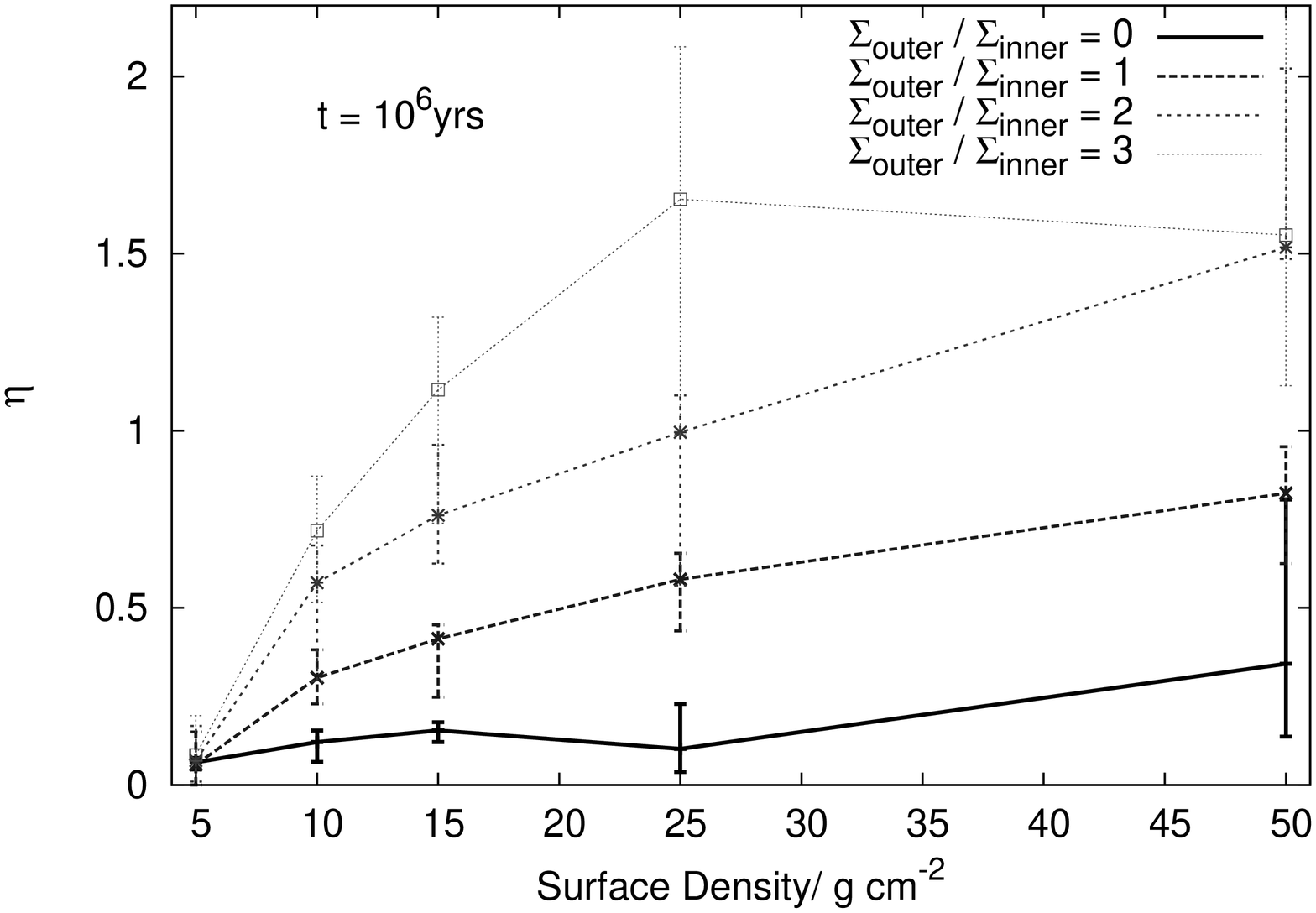,angle=-90,width=\columnwidth}}
\centerline{\psfig{figure=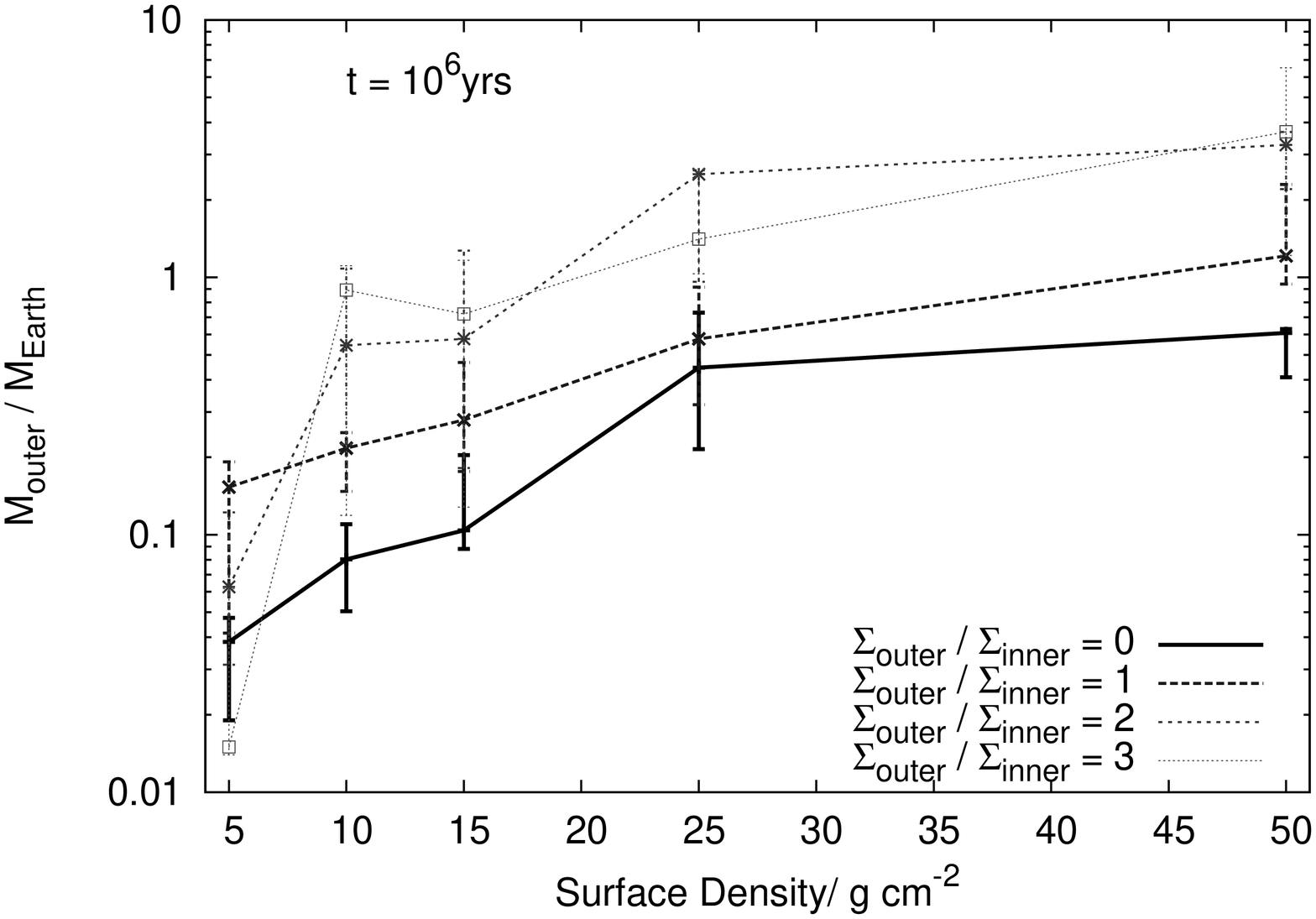,angle=-90,width=\columnwidth}}
\caption{Variation in scattering driven migration (top, $\eta = (a_{outer,final} - a_{outer,initial})/a_{outer,initial}$) and mass (bottom) due to changes in surface density. The plots are (in order dark-to-light), $\Sigma_{outer} / \Sigma_{inner} = 0$; $\Sigma_{outer} / \Sigma_{inner} = 1$;  $\Sigma_{outer} / \Sigma_{inner} = 2$; $\Sigma_{outer} / \Sigma_{inner} = 3$. Increases in surface density above the MMNM ($10\textrm{g cm}^{-2}$) cause significantly increased migration. Plotted are the median values at $10^6\,yrs$ (10 simulations for each point), with the error bars giving the upper and lower quartiles.}
\label{FIG:surf_den_A}
\end{figure}

We plot in Fig \ref{FIG:surf_den_A} the results of the surface density variations after $10^6$ years. We see that in general, moving to a higher absolute surface density increases the degree of outward migration, as does a move towards a higher surface density ratio: at $\Sigma_0 = 10 \textrm{g cm}^{-2}$ the relative change in outer semi-major axis, $\eta \approx 0.3$, but for $\Sigma_0 = 50 \textrm{g cm}^{-2}$, this increases to $\eta = 0.79_{-0.24}^{+0.14}$. Similarly, the mass of the outer-most body also increases significantly as the surface density and the surface density ratio are increased.

In the extreme case that the number of planetesimals in the outer zone goes to zero, then the outward migration of the embryos from the inner zone is heavily suppressed, confirming the requirement that there be an external population of some kind to enable scattering-driven migration to take place. 

%

We note at this point that as the surface density is changed between simulations, the initial number of embryos also varies (because the isolation mass and semi-major axis separation are functions of surface density - see Eqn \ref{EQN:mass}), changing from 80 bodies when $\Sigma =5$ to 25 bodies when $\Sigma = 50$. This change in the number of bodies does itself have consequences for the scattering rate (see \S \ref{Number}), so we re-ran the simulations with the initial number of embryos artificially constrained to be a constant (56), and found that the change in the initial number of bodies (e.g. going from 25 to 56 bodies at $\Sigma = 50$) caused only a small modification to the migration and mass growth compared to that resulting from the overall surface density variation.

Previous studies by \citet{Guedes08} which looked at the evolution of embryos \emph{after} the runaway-growth phase used a shallower surface density profile ($\Sigma \propto a^{-1}$) than that used in the majority of this work ($\Sigma \propto a^{-3/2}$). \citet{Thebault09} established that such a shallower profile did \emph{not} help to extend the inner zone which is suitable for runaway growth. For the sake of completion, we investigate in set D whether a change in the surface density slope would have any effect on the outward migration in our n-body simulations. Changing the surface density profile obviously changes the relative masses at given points, so we set up simulation set D so that the absolute normalisations at $1\,AU$ were \emph{different}, changing them so that the total mass in the \emph{inner} zone for each simulation was identical ($\sim 2\Mearth$). We found as expected that shallower slopes, which effectively increase the mass ratio between the outer and inner zones, increase the degree of outward migration, so that (at $10^6\,yrs$), the relative outward migration, $\eta = 1.04_{-0.31}^{+0.15}$ for a completely flat ($\Sigma \propto a^0$) simulation (compared to $\eta = 0.29$ for $\Sigma \propto a^{-3/2}$).

\subsection{Position of the Zone Boundary}\label{Boundary}
The simulations thus far have assumed that the boundary, $a_{crit}$, between the inner and outer zones lies at $0.7\,AU$, as appropriate for the nominal case of $\alpha$ Cen-B. However, the position of $a_{crit}$ is expected to vary significantly for a number of reasons: (i) In \citet{Thebault09}, $0.7\,AU$ was in fact the position that would be expected for a rather massive disk in which embryo formation extended to rather greater radii than would expected in a MMNM disk; (ii) Variations in inclination \citep[see][]{Xie09} have been found to change the size of the region over which runaway growth is suppressed; and (iii) Differing relative binary parameters (mass ratio, separation and eccentricity) would all serve to change the location of $a_{crit}$. We therefore generalise our investigation to look at how our results will scale for other systems in which $a_{crit}$ lies at a different distance from the central star. 

We note that as we keep the total mass and extent of our disk constant, moving $a_{crit}$ in (out) has the effect of decreasing (increasing) the total number of isolation mass embryos that are simulated within the inner zone. We thus always calculate the $\eta$-migration parameter by focusing on the outer-most embryo at any given time.

\begin{figure}
  \centerline{\psfig{figure=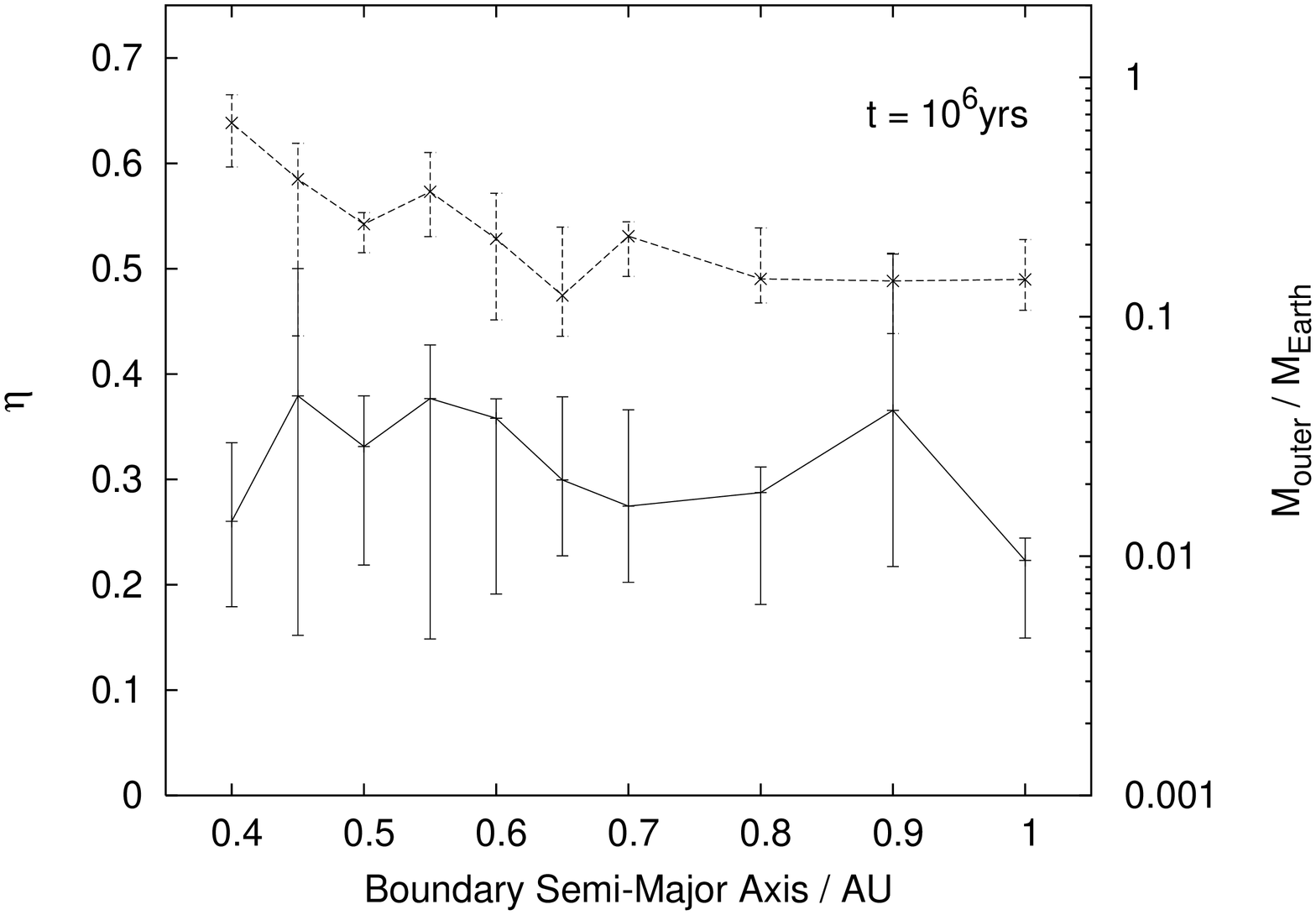,angle=-90,width=\columnwidth}}
  \centerline{\psfig{figure=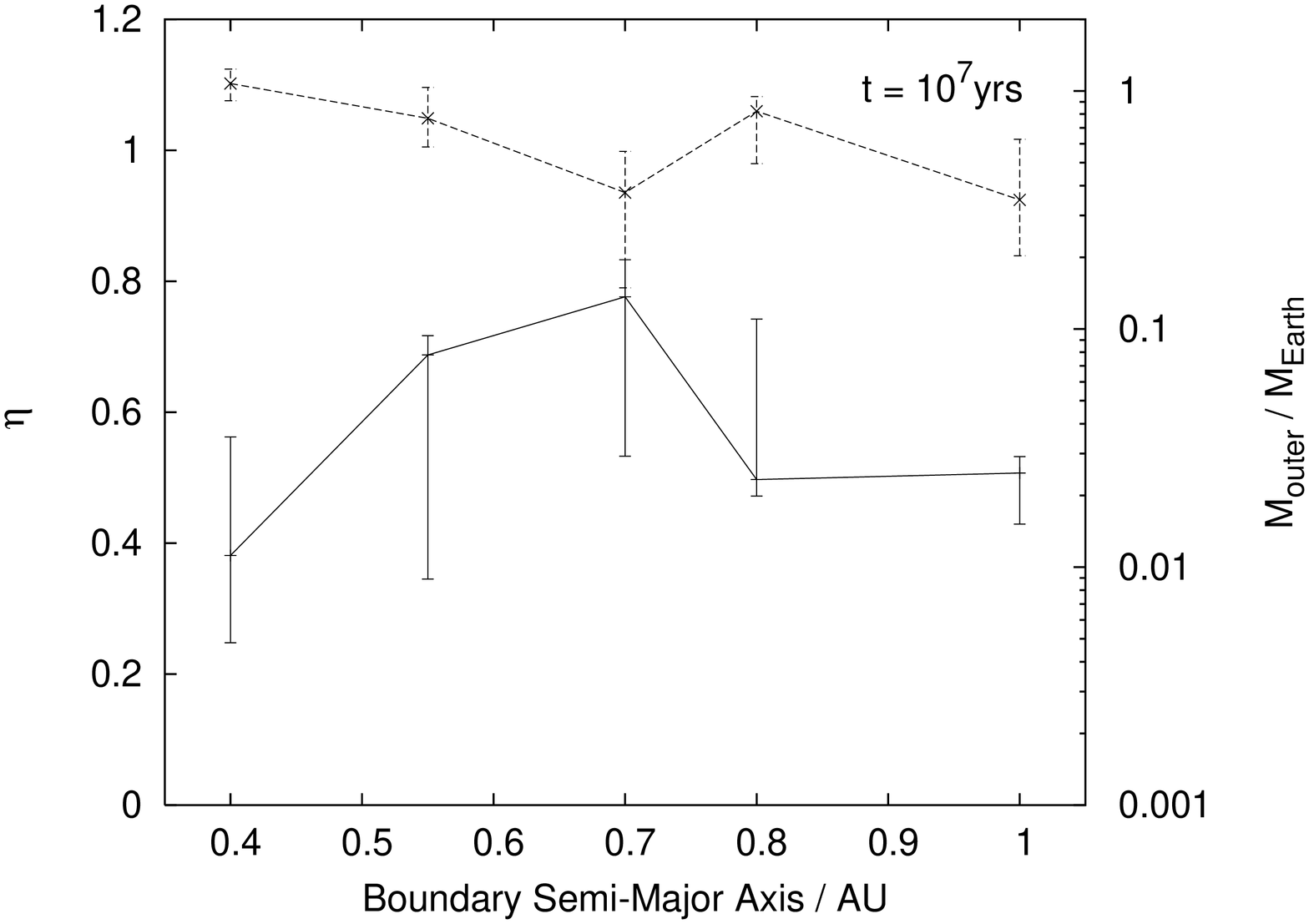,angle=-90,width=\columnwidth}}
  \caption{Variation in relative migration, $\eta = (a_{outer,final} - a_{outer,initial})/a_{outer,initial}$ (solid line, left-hand axis) and mass (dashed line, right-hand axis) of the outer-most embryo as a function of the boundary radius, $a_{crit}$, between inner and outer zone. Each point gives the median value from 10 simulations, with the error bars giving the upper and lower quartiles. Top: $10^6yrs$ - The \emph{relative} migration is approximately constant at $\sim 0.3$ suggesting that the relative change in semi-major axis of the outer-most embryo will be effectively independent of the precise initial value of $a_{crit}$. The mass of this outer-most embryo shows a gentle decline as a function of $a_{crit}$. Bottom: $10^7yrs$ - The variation in the relative migration, $\eta$, is large but again does not show an obvious trend as a function of $a_{crit}$.}
  \label{FIG:F}
\end{figure}

In set E we look at the degree of outward migration that results for disks in which $a_{crit}$ varies between 0.4 and $1.0\,AU$. In \S \ref{Typical} we found that for our fiducial MMNM with $a_{crit} = 0.7\,AU$, that the average relative change in semi-major axis of the outer-most embryo was $\eta = 0.3$. Varying $a_{crit}$ in Fig \ref{FIG:F} reveals that at $10^6$ years $\eta$ is largely unaffected by changes in $a_{crit}$: there is some scatter in the results obtained, but there is no obvious trend in the degree of migration, suggesting that the \emph{relative change} in semi-major axis of the outermost embryo will effectively be independent of the precise location of $a_{crit}$. However, we note that the mass of the outer-most embryo declines as a function of $a_{crit}$, as the embryos collide with a smaller number of planetesimals at larger semi-major axes.

When we extend the simulations on to $10^7$ years, as is depicted in the lower panel of Fig \ref{FIG:F}, we find that the situation is perhaps slightly more complicated than initially thought, as the value of $\eta$ at this time is now seen to vary more widely with $a_{crit}$. One could argue that there is perhaps an approximate peak in relative migration for $a_{crit} \sim 0.7\,AU$, but the variation in simulation results is so large (especially for low $a_{crit}$) that it seems more sensible to take the view that the results are consistent with the relative migration being approximately constant at $\eta \sim 0.6$.

Finally, we now consider what happens if we dispense with $a_{crit}$ completely, and simply consider a MMNM disk in which growth to isolation has \emph{not} been suppressed for any $a$, i.e. the situation as might be envisaged around a single (non-binary) star. We would expect it to have a chain of isolation mass embryos extending across the entire terrestrial region. We model such a disk, simulating the evolution of a chain of embryos initially spread between 0.2 and $1.5\,AU$. As might be anticipated, we find that there is a considerable degree of orbit crossing, scattering and collision amongst the embryos. 

If we consider just those embryos which start with $a < 0.7\,AU$ we find that after $10^6yrs$ ($10^7\,yrs$), the maximum semi-major axis to which this material has scattered / migrated, is such that $\eta = 0.29_{-0.13}^{+0.11}$ ($\eta = 0.83_{-0.94}^{+0.13}$). Thus there is a rather similar degree of ``mixing'' of material from $a < a_{crit}$ to $a > a_{crit}$ as was seen in set A in which the initial embryo growth beyond $a_{crit}$ was suppressed. But it should be noted that these simulations with no $a_{crit}$ show much more variability than those in set A, with a third of the simulations showing \emph{no} spread at all beyond $0.7\,AU$, where-as \emph{all} of the simulations in set A migrated to at least $1.0\,AU$ by $10^7\,yrs$ (with the average being $1.25\,AU$, $\eta = 0.8$).

\subsection{Varying the Composition of the Inner System}\label{Additional}
\subsubsection{Number and eccentricity of the inner embryo population.}\label{Number}
In set F we vary the initial content of the inner zone, keeping the total mass and surface density profile the same as in the fiducial case, but we now vary the number and individual masses of the embryos, and then examine the effect of varying their initial eccentricity. If the number of embryos in the inner zone is substantially lower than the 56 which results from assuming $\Delta a = 7r_H$ (i.e. if we make $K>7$), then this would be more consistent with a set of initial conditions occurring a substantial time period \emph{after} the gas disk has dissipated and the subsequent stage of chaotic collision has proceeded for some time, resulting in more widely spaced (and more massive) embryos.

As the initial number of embryos is decreased to 25, the amount of outward migration is \emph{decreased}, with the outer-most embryo at $10^6yrs$ having migrated such that $\eta = 0.21_{-0.08}^{+0.06}$ but the mass of the outer-most body increases to $\sim 0.3\Mearth$, I.e. larger bodies have less chance of being scattered / migrating outwards. By the time the initial number of embryos is lowered to 10, the embryos are so widely separated, and the interaction timescales so long \citep{Zhou07}, that the bodies have little chance of being scattered away from their initial locations.

\begin{figure}
  \centerline{\psfig{figure=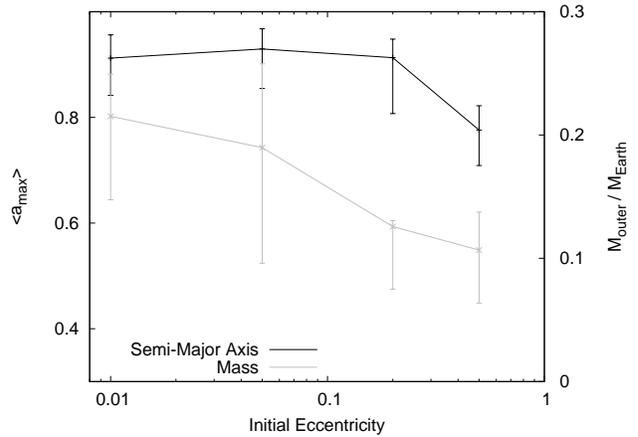,angle=-90,width=\columnwidth}}
  \caption{Variation in the semi-major axis (black line, left-hand axis) and mass (grey line, right-hand axis) as a result of variations in the initial eccentricity of the embryo population. Results are measured at $t=10^6\,yrs$. It can be seen that as long as the eccentricity is ``reasonable'', i.e. less than 0.5, than there is no real trend for the maximum semi-major axis of the outer planet to vary with eccentricity. However, there does seem to be a trend for the mass to decrease with eccentricity, such that $M \propto -\log (e)$.}
  \label{FIG:ECCDEP}
\end{figure}
%
In a scenario such as this (in which the embryos are more widely spaced as a result of an assumed period of scattering), we would expect that the eccentricities of the embryos would be substantially higher, approaching $e\sim 0.2$ as the number of bodies drops into single figures \citep{Chambers01}. We investigate this, but find from further simulation work (see \Fig{FIG:ECCDEP}) that the initial eccentricity is largely unimportant in determining the degree of migration, as long as it is not above an improbably large value ($\sim 0.5$). Interestingly however, we find that the mass of the outer object \emph{is} influenced by the initial eccentricity, with initial eccentricities of $e=0.01$ ($e=0.2$) resulting in masses for the migrating embryos of $0.2\Mearth$ ($\sim 0.1\Mearth$), I.e. giving an approximate linear fit:
\beq
M_{Outer,\,Final} (\textrm{t=}10^6\textrm{ yrs}) &\approx & 0.08 - 0.06 \log_{10}\left(e_{Initial}\right), \label{ME}
\eeq
where we stress that this result is applicable to measurements taken after $10^6$ years of the evolution of simulations of a $10\, g\, cm^{-2}$ surface density disk, in which isolation-mass embryos were initially confined to a region $a < 0.7\, AU$.

\subsubsection{ Introduction of Planetesimals to the Inner System.}
It is possible that at the time of disk dissipation, the inner zone may contain a mixed population of massive embryos and less-massive planetesimals that have not yet been accreted. In set G we vary the fraction of the inner zone which is composed of embryos, varying the fraction, $f$, from the the value of 1 used in previous sets down to 0.1. 

If we implement this variation by keeping the surface density, mass and number of embryos constant but \emph{add planetesimals to the inner zone} (and also add planetesimals to the outer zone proportionately), then we have increased the overall surface density, and qualitatively the results are as might be expected from \S \ref{Results_Sigma}, with both migration rates and growth rates increasing. We find at $10^6\,yrs$, that for $f=0.1$ (i.e. meaning the total surface density is ten times higher), $\eta = 1.4_{-0.35}^{+0.10}$.

If we implement the variation by keeping the overall surface density constant, requiring the addition of planetesimal mass to the inner zone and the subtraction of embryonic mass (and leave the outer zone unaltered), then we find little difference between the $f = 0.1$ and $f=1$ cases, with both having $\eta \sim 0.3$ at $10^6\,yrs$. The planetesimals do not seem to have any significant damping effect.

\subsection{Simulation Approximations}\label{Approximations}
\subsubsection{Planetesimals}
Ideally we would have liked to perform our simulations using a vast number of very small planetesimals in the outer region, fully interacting with the embryo population as they orbited the central star under the influence of the perturbing binary. However, for the sake of computational efficiency and practicality we simulated only the planetary embryos as fully interacting particles, modelling the planetesimals as test particles with an associated mass. This means that gravitational embryo-embryo and embryo-planetesimal interactions took place in the simulations, but planetesimal-planetesimal interactions were neglected. We also used only a relatively small number of massive planetesimals ($\sim 100$). In simulation sets H and I, we explicitly investigate the implications of these approximations by conducting a limited number of simulations in which the number of planetesimals is increased and / or the planetesimals are modelled as fully-interacting n-body particles.

Our fiducial set A simulations in which the mass of planetesimals in the outer disk is distributed amongst 100 planetesimals, found that the outermost embryo migrated outwards such that $\eta = 0.29$ after $10^6 yrs$. When we distribute the planetesimal mass between 1,000 bodies (set H), we find that the outermost body has migrated by the same amount (again $\eta = 0.31_{-0.07}^{+0.06}$) in the same time period, suggesting that the degree and rate of migration are essentially unaffected when using planetesimal numbers $\geq 100$.

Additionally, we find that it does not seem to make a significant difference to the migration history if we use a fully-interacting model for the planetesimals, rather than modelling them as massive test particles. Switching between set A (no planetesimal-planetesimal scattering) and set I (planetesimal-planetesimal scattering allowed) resulted in the average maximum migration distance of the outermost embryo being almost identical ($0.89\,AU$). We therefore feel confident that our approximation of using a relatively small number of massive test-particles provides sufficient accuracy to allow us some insight into the behaviour of such systems.

\subsubsection{Binary Companion}\label{Binary Companion}
As a final point, we turn to consider in greater detail whether it was reasonable to have ignored the presence of the binary perturber in our simulations. We now explicitly include the binary companion in our simulation set J, starting the simulations with a $1.1\Msun$ mass companion at $18\,AU$ with $e=0.5$ (i.e. using the parameters of the $\alpha$ Cen system, assuming the embryos are in orbit around $\alpha$ Cen-B). We also slightly alter the initial conditions for the embryo and planetesimal disk, taking into account the likely forced eccentricity due to the binary object. Using the standard first order expansion in the disturbing function \citep{Heppenheimer78,MurrayandDermott}, we find that $e_{forced} \approx 0.035 a$. Other  than this, the simulations are kept the same as Set A, with the same number of embryos and planetesimals, and the planetesimals performing the same scattering role. 

\begin{figure}
  \centerline{\psfig{figure=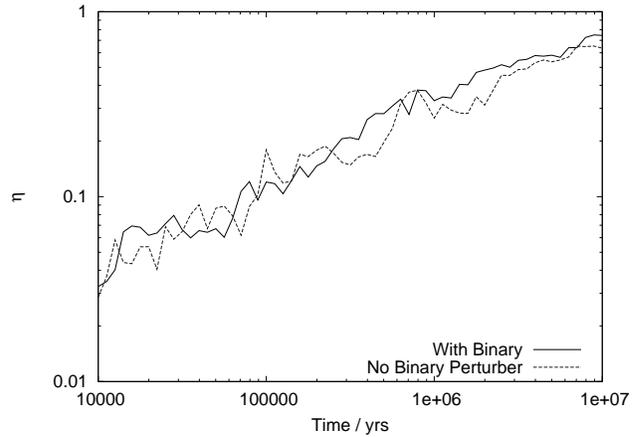,angle=-90,width=\columnwidth}}
  \caption{We compare the evolution of the $\eta$-parameter (see \Eqn{EQN_ETA_DEFN}) in simulation set J (black line, includes binary perturber) with the results from set A (grey line, no binary perturber). The behaviour of the simulations is almost identical, suggesting that the binary companion does not significantly change the outward migration.}
  \label{FIG:31}
\end{figure}

A comparison between the results of this simulation set (J) and the results of our fiducial set (A) is provided in Fig. \ref{FIG:31}, demonstrating that there is very little difference between the two results, suggesting that the binary companion is not important in dictating the degree of scattering-driven migration.

\begin{figure}
  \centerline{\psfig{figure=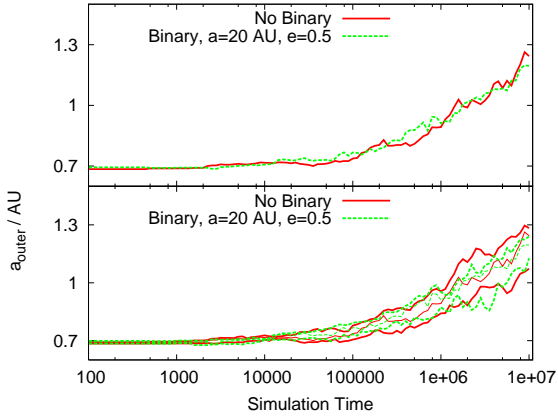,angle=-90,width=\columnwidth}}
  \caption{Variation in average maximum embryo semi-major axis for simulations with and without a binary perturber. Green plots include the perturber and a realistic forced eccentricity, Red plots do not have a perturber present in the simulation. The top panel displays the median semi-major axis, whilst the bottom panel also displays the evolution of the upper and lower quartile values from the 10 simulations conducted. The migration distance is essentially unchanged by the presence of the perturber.}
  \label{FIG:21}
\end{figure}

\begin{figure}
  \centerline{\psfig{figure=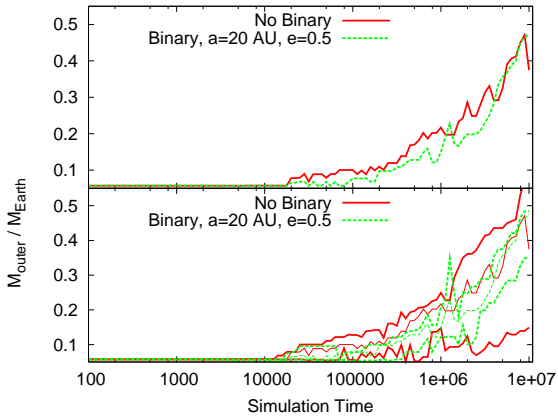,angle=-90,width=\columnwidth}}
  \caption{Variation in mass of the outer-most embryo for simulations with and without a binary perturber. Green plots include the perturber and a realistic forced eccentricity, Red plots do not have a perturber present in the simulation. The top panel displays the median mass, whilst the bottom panel also displays the evolution of the upper and lower quartile values from the 10 simulations conducted. The median mass of the outer-most embryo is also unaltered by the presence of the binary, although it should be noted that the simulations \emph{without} the binary have a larger spread in masses. }
  \label{FIG:22}
\end{figure}

Investigating further, we plot in Fig \ref{FIG:21} the semi-major axis of the outermost embryo, demonstrating that its median value, as well as the upper and lower quartile spreads are essentially unchanged by the presence of the binary companion. The \emph{mass} of the outer-most body in Fig \ref{FIG:22} is also approximately the same for the more realistic simulation involving the binary companion plus initial forced eccentricity, but it should be noted that the simulations \emph{with} the the binary companion (set J) have a much smaller spread in the upper and lower quartiles for the mass.


\begin{figure}
  \centerline{\psfig{figure=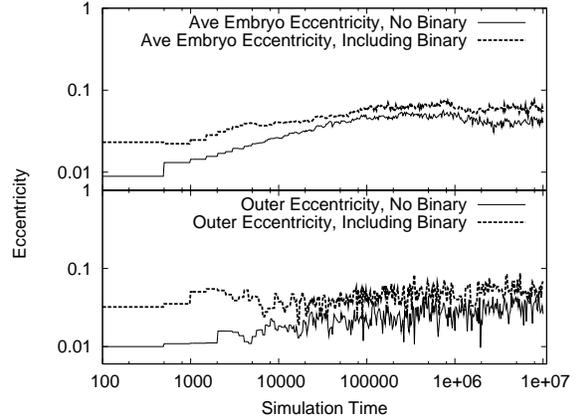,angle=-90,width=\columnwidth}}
  \caption{We plot the evolution in the eccentricity of the embryos in simulation sets A and J, i.e. we compare the eccentricity evolution excluding and including the presence of the perturbing binary. In the top panel we show the average eccentricity evolution of the entire embryo population, whilst in the bottom panel we focus on the eccentricity of the outer-most embryo. In both panels the thin lines give the results of set A in which no binary companion is present , whilst the thick line gives the results for set J in which the binary companion is explicitly simulated. We find that the behaviour of both measures is rather similar, with the initial forced eccentricity being obvious in both panels, raising the eccentricities above the cases with no binary perturber. However, in both cases the difference between set A and set J diminishes with time as the embryos stir themselves and increase their eccentricity.}
  \label{FIG:ECC_COMP}
\end{figure}

Finally, we examine the eccentricities of the bodies in Fig \ref{FIG:ECC_COMP}, finding that both the eccentricity of the outer-most embryo \emph{and} the average eccentricity of all the embryos evolve in rather similar manners: the simulations with a binary perturber start with a larger (forced) eccentricity, but over time this difference becomes of less significance as the embryos in all simulations start to stir themselves to similar levels. We note that  in \S \ref{Number} the accreted mass was seen to reduce slightly for simulations with higher eccentricity: if we substitute the comparatively small forced eccentricity (0.035) into \Eqn{ME}, then we find that the expected reduction would be of only a few percent. 

We take this opportunity to reiterate that, even though the binary companion was omitted from the majority of the n-body simulations, its suppression effect on runaway growth has of course been included via $a_{crit}$ and the attendant lack of embryos for $a > a_{crit}$.

\section{Conclusion}
We have considered the formation and migration of protoplanetary embryos in disks around the stars in binary systems, in which the initial stages of embryo formation have been curtailed beyond some critical semi-major axis, $a_{crit}$. We find that: 
\begin{itemize}
\item For a fiducial MMNM disk, over a period of $10^6yrs$ ($10^7yrs$), systems will on average exhibit outward migration such that they contain embryos which have migrated from $0.7\,AU$ to $0.9\,AU$ ($1.2\,AU$), i.e. they have experienced a relative change in semi-major axis of $\eta = 0.3$ ($0.8$), with the average outer-most body having a mass of $0.2\Mearth$ ($0.4\Mearth$) (see Fig \ref{FIG:Fiducial}). 
\item The migration starts to plateau by the $10^7\,yr$ mark in the simulations, suggesting that little significant further migration is to be expected beyond this point (see Fig \ref{FIG:Fiducial}).
\item Higher surface density disks increase the rate and degree of migration as well as increasing the mass of the migrating body (see Fig \ref{FIG:surf_den_A}).
\item The position of the zone boundary ($a_{crit}$, outside which runaway growth is suppressed), is unimportant in deciding the degree of relative migration, $\eta$, that takes place in the system: whilst there is significant scatter in the results obtained, they are consistent with $\eta$ being independent of $a_{crit}$ at a given time (see Fig \ref{FIG:F}). 
\item The presence of the binary companion in the system, whilst important in initially determining the cut-off radius for embryo growth, does \emph{not} significantly impact the subsequent outward migration rate of the embryos, nor their final masses (see Figs \ref{FIG:31}, \ref{FIG:21} and \ref{FIG:22}).
\end{itemize}

We suggest that tight binary systems, where runaway embryo growth is initially confined to $a < a_{crit}$, may very generally experience outward migration that is sufficient to move these planetary embryos to regions $a > a_{crit}$. This allows the possibility that the habitable zone can be populated with detectable, terrestrial mass planets, even if $a_{crit}$ is significantly interior to the HZ.

\section*{Acknowledgments}
The authors acknowledge the High-Performance Computing Centre of the University of Cambridge for providing computational resources and support. The authors are also grateful to our referee who greatly helped improving the manuscript. MJP thanks the UK PPARC/STFC for a research studentship.


\label{lastpage}
\end{document}